\documentclass{aa}  
\usepackage{graphicx}
\usepackage{txfonts}
\usepackage{xcolor}
\usepackage{hyperref}
\hypersetup{colorlinks, linkcolor=blue, citecolor=blue, urlcolor=blue}
\usepackage{amsmath}
\usepackage{ulem}
\usepackage{soul}
\usepackage{enumitem}% http://ctan.org/pkg/enumitem
\newcommand{\muvec}{\mbox{\boldmath $\mu$}}

\newcommand{\te}{t_{\rm E}}
\newcommand{\thetae}{\theta_{\rm E}}

\newcommand{\pie}{\pi_{\rm E}}

\newcommand{\dl}{D_{\rm L}}
\newcommand{\ds}{D_{\rm S}}

\definecolor{brown}{rgb}{0.59, 0.29, 0.0}
\definecolor{darkgreen}{rgb}{0.0, 0.42, 0.24}
\definecolor{darkblue}{rgb}{0.01, 0.31, 0.59}
\definecolor{darkblue}{rgb}{0.0, 0.25, 0.42}
\definecolor{blue}{rgb}{0.0,0.0,1.0}
\definecolor{green}{rgb}{0.0,1.0,0.0}

\def\eqalign#1{\null\,\vcenter{\openup\jot
        \ialign{\strut\hfil$\displaystyle{##}$&$
        \displaystyle{{}##}$\hfil \crcr#1\crcr}}\,}
\begin{document}

\title{KMT-2024-BLG-1044L: A sub-Uranus microlensing planet around a host at the star--brown dwarf mass boundary} 
\titlerunning{KMT-2024-BLG-1044L}

\author{
% leading author -----------------------------
     Cheongho~Han\inst{\ref{inst1}} 
\and Yoon-Hyun~Ryu\inst{\ref{inst2}}
\and Chung-Uk~Lee\inst{\ref{inst2}, \ref{inst9}} 
\and Andrew~Gould\inst{\ref{inst3},\ref{inst4}}      
% KMTNet ---------------------------
\and Michael~D.~Albrow\inst{\ref{inst5}}   
\and Sun-Ju~Chung\inst{\ref{inst2}}      
\and Kyu-Ha~Hwang\inst{\ref{inst2}} 
\and Youn~Kil~Jung\inst{\ref{inst2}} 
\and Yossi~Shvartzvald\inst{\ref{inst6}}   
\and In-Gu~Shin\inst{\ref{inst7}} 
\and Jennifer~C.~Yee\inst{\ref{inst7}}   
\and Hongjing~Yang\inst{\ref{inst8}}     
\and Weicheng~Zang\inst{\ref{inst7},\ref{inst8}}     
\and Doeon~Kim\inst{\ref{inst1}}
\and Dong-Jin~Kim\inst{\ref{inst2}} 
\and Byeong-Gon~Park\inst{\ref{inst2}} 
\and Richard~W.~Pogge\inst{\ref{inst4}}
\\
(The KMTNet Collaboration)
}

\institute{
      Department of Physics, Chungbuk National University, Cheongju 28644, Republic of Korea\label{inst1}                             % (1)
\and  Korea Astronomy and Space Science Institute, Daejon 34055, Republic of Korea\label{inst2}                                       % (2)
\and  Max Planck Institute for Astronomy, K\"onigstuhl 17, D-69117 Heidelberg, Germany\label{inst3}                                   % (3)
\and  Department of Astronomy, The Ohio State University, 140 W. 18th Ave., Columbus, OH 43210, USA\label{inst4}                      % (4)
\and  University of Canterbury, Department of Physics and Astronomy, Private Bag 4800, Christchurch 8020, New Zealand\label{inst5}    % (5)
\and  Department of Particle Physics and Astrophysics, Weizmann Institute of Science, Rehovot 76100, Israel\label{inst6}              % (6)
\and  Center for Astrophysics $|$ Harvard \& Smithsonian 60 Garden St., Cambridge, MA 02138, USA\label{inst7}                         % (7)
\and  Department of Astronomy and Tsinghua Centre for Astrophysics, Tsinghua University, Beijing 100084, China\label{inst8}           % (8)
\and  Corrresponding author\label{inst9}                                                                                              % (9)  
}
\date{Received ; accepted}

% \abstract{}{}{}{}{} 
% 5 {} token are mandatory
\abstract
{}
{
We analysed microlensing data to uncover the nature of the anomaly that appeared near the peak 
of the short-timescale microlensing event KMT-2024-BLG-1044. Despite the anomaly's brief duration 
of less than a day, it was densely observed through high-cadence monitoring conducted by the KMTNet 
survey.
}
{
Detailed modelling of the light curve confirmed the planetary origin of the anomaly and revealed 
two possible solutions, due to an inner--outer degeneracy.  The two solutions provide different 
measured planet parameters: 
$(s, q)_{\rm inner} = [1.0883 \pm 0.0027, (3.125 \pm 0.248)\times 10^{-4}]$ for the inner solutions and
$(s, q)_{\rm outer} = [1.0327 \pm 0.0054, (3.350 \pm 0.316)\times 10^{-4}]$ for the outer solutions.
}
{
Using Bayesian analysis with constraints provided by the short event timescale ($t_{\rm E} \sim 9.1$~day) 
and the small angular Einstein radius ($\theta_{\rm E}\sim 0.16$~mas for the inner solution and 
$\sim 0.10$~mas for the outer solutio), we determined that the lens is a planetary system consisting of 
a host near the boundary between a star and a brown dwarf and a planet with a mass lower than that 
of Uranus. The discovery of the planetary system highlights the crucial role of the 
microlensing technique in detecting planets that orbit substellar brown dwarfs or very low-mass stars.
}
{}


\maketitle

\section{Introduction}\label{sec:one}

\citet{Mao1991} and \citet{Gould1992} first pointed out that microlensing could be used
as a tool for detecting extrasolar planets.  Experimental searches for planets began in the
mid-1990s, and the first microlensing planet was detected in 2003 by \citet{Bond2004}. To 
date, 221 microlensing planets have been reported, according to the NASA Exoplanet 
Archive\footnote{\tt https://exoplanetarchive.ipac.caltech.edu/}.  Currently, microlensing 
ranks as the third most prolific planet detection method, behind the transit and radial 
velocity methods.

Planetary signals in lensing light curves are 
typically brief. Consequently, detecting these short-lived signals in microlensing light 
curves necessitates dense coverage of lensing events.  To address this need, early planetary 
searches combined large-scale survey programmes, such as the Optical Gravitational Lensing 
Experiment \citep[OGLE;][]{Udalski1994}, the Massive Astrophysical Compact Halo Object 
survey \citep[MACHO;][]{Alcock1993}, and Microlensing Observations in Astrophysics 
\citep[MOA;][]{Bond2001}, with dedicated follow-up observation networks. These follow-up 
networks included the Galactic Microlensing Alerts Network \citep[GMAN;][]{Alcock1997}, the 
Probing Lensing Anomalies NETwork \citep[PLANET;][]{Albrow1998}, the Microlensing Follow-Up 
Network \citep[$\mu$FUN;][]{Gould2006}, and RoboNet \citep{Tsapras2003}.  In this mode of 
observations, survey experiments monitored a wide region of the Galactic bulge to detect 
lensing events, while follow-up groups utilised multiple narrow-field telescopes to conduct 
dense observations of the events identified by the surveys.  This survey+follow-up strategy only 
allowed for dense observations of a limited number of events, resulting in fewer than 
five planetary detections per year before 2010.  From 2010 to 2015, the number of detections 
nearly doubled with a moderate modification to the observational strategy. In the modified 
mode, follow-up observations were mainly focused on the peak phases of high-magnification 
events, for which the probability of planet detection is high.  Another strategy employed 
during this period was the auto-follow-up mode.  In this approach, when a survey detects a 
potential anomalous lensing event in progress, an alert is triggered to promptly initiate 
follow-up observations \citep{Suzuki2016}.  A dramatic increase in planet detections occurred 
with the commencement of high-cadence surveys using multiple telescopes equipped with very 
wide-field cameras. These surveys significantly increased the number of lensing event detections 
from several dozen per year to over 3000. With the ability to densely monitor all detected 
lensing events, the number of planet detections also dramatically increased, and currently 
an average of 25 planets are being reported annually.

The primary significance of the microlensing technique is its ability to detect planetary systems
that might be difficult to identify using other observational methods. While transit and radial
velocity methods are sensitive to planets that are close to their host stars, the microlensing
method is adept at detecting planets in wide orbits.  Additionally, microlensing can identify 
distant planetary systems located in both the disc and bulge regions of the Galaxy.  While planets 
are typically detected through their effect on their host stars, the microlensing effect can be 
caused by the planet's own gravity, enabling the detection of free-floating planets.\footnote{
Currently, nine free-floating planet candidates have been reported \citep{Mroz2018, Mroz2019, 
Mroz2020a, Mroz2020b, Kim2021a, Ryu2021a, Koshimoto2023, Jung2024}.} Furthermore, because 
microlensing relies on the gravitational influence of a lensing object, it possesses a unique 
capability to detect planets belonging to dark or very faint celestial bodies, such as brown 
dwarfs (BDs), very faint stars, and stellar remnants.  A comprehensive discussion of the various 
advantages of the microlensing method is provided in the review by \citet{Gaudi2012}.  This 
capability makes microlensing an invaluable tool for obtaining a comprehensive view of the 
demographics of planetary systems throughout the Galaxy.

In this study we report the discovery of a low-mass planet orbiting a very low-mass host. The
planetary system was discovered from the analysis of a brief anomaly signal in the light
curve of the short-timescale lensing event KMT-2024-BLG-1044. Based on the constraints provided
by the event timescale together with the angular size of the Einstein ring, it was estimated that
the planet has a mass lower than that of Uranus and that the host has a mass at the
boundary between BDs and low-mass stars.

\section{Observations and data}\label{sec:two}

The lensing event KMT-2024-BLG-1044 was detected by the Korea Microlensing Telescope Network 
\citep[KMTNet;][]{Kim2016} survey during the 2024 season using the EventFinder algorithm 
\citep{Kim2018}. The KMTNet survey operates three identical telescopes strategically positioned 
in the Southern Hemisphere: the Cerro Tololo Inter-American Observatory in Chile (KMTC), the 
South African Astronomical Observatory in South Africa (KMTS), and the Siding Spring Observatory 
in Australia (KMTA).  On May 20, 2024 (HJD$^\prime =  450$), when the event was detected, the 
source flux had just passed its peak magnification. Here, ${\rm HJD}^\prime \equiv {\rm HJD} - 
2460000$ represents a shortened heliocentric Julian date. The magnitude of the source before 
the onset of lensing magnification was $I_{\rm base}=18.85$, and it reached $I_{\rm peak}=17.82$ 
at its highest point. The equatorial and Galactic coordinates of the source are $({\rm RA}, 
{\rm Dec})_{\rm J2000} = $(18:06:15.36, -27:35:33.29) and $(l, b) = (3^\circ\hskip-2pt .4484, 
-3^\circ\hskip-2pt .2509)$, respectively.  The extinction towards the field was $A_I = 0.98$.  
The source is located in the overlapping region of the KMTNet prime fields BLG03 and BLG43, 
towards which KMTNet observations were most frequently conducted. The observation cadence was 
0.5~hours for each field and 0.25~hours for the combined field. Thanks to this high cadence, 
the light curve was densely covered despite the relatively short duration of the event.

\begin{figure}[t]
\includegraphics[width=\columnwidth]{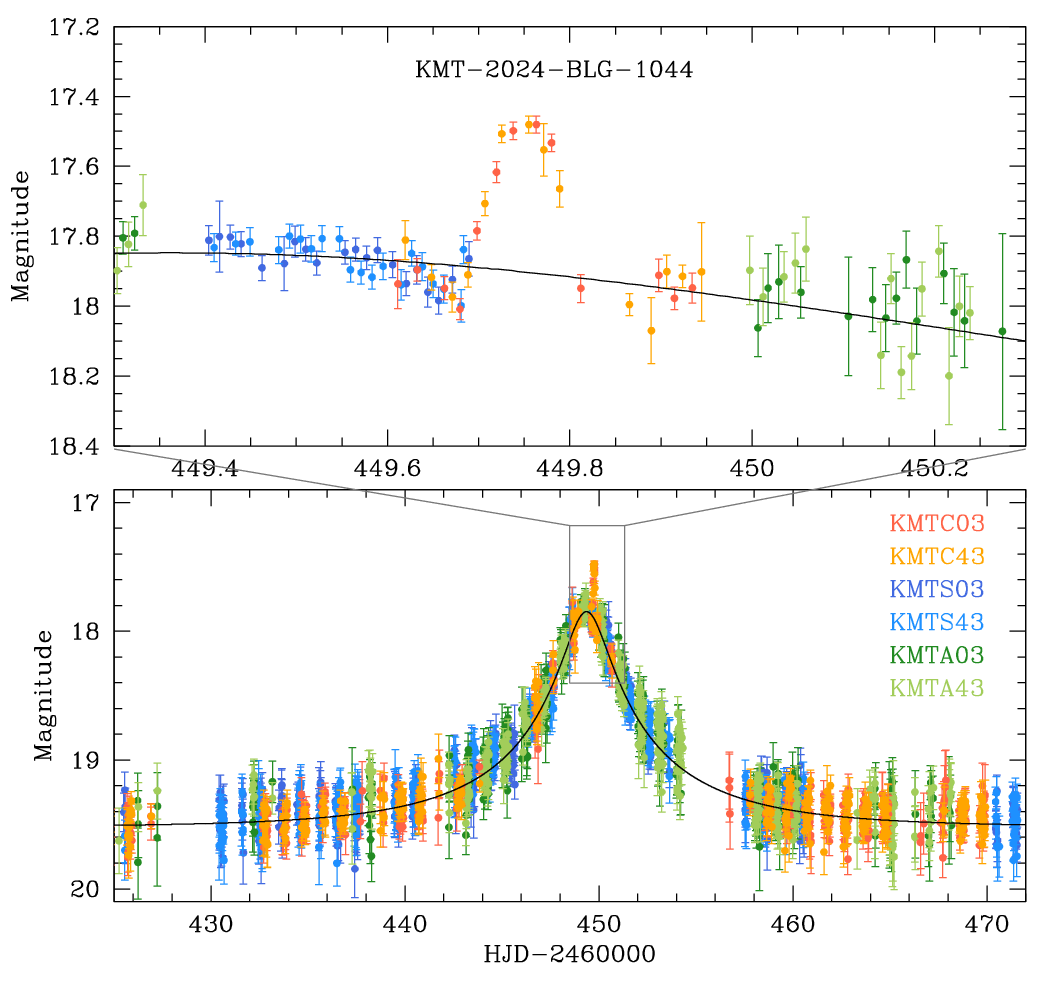}
\caption{
Light curve of the lensing event KMT-2024-BLG-1044.  Lower panel: Full view of the 
light curve. Upper panel: Zoomed-in view of the peak region around the anomaly. 
The solid curve drawn over the data points is a single-lens single-source model obtained by fitting 
the data excluding those around the anomaly. The colours of the data points correspond to those of the datasets marked in the legend.      
}
\label{fig:one}
\end{figure}

Figure~\ref{fig:one} displays the light curve of the lensing event, compiled from all KMTNet 
datasets.  At first glance, the light curve appears to be that of a typical single-lens 
single-source (1L1S) event with a smooth and symmetrical shape. However, upon closer examination, 
a brief anomaly near the peak of the light curve was identified. This anomaly persisted for 
less than a day, occurring between ${\rm HJD}^\prime = 449.6$ and 449.9. The upper panel 
provides a magnified view of this anomaly.  The primary portion of the anomaly showed a 
positive deviation from the 1L1S model.  Additionally, there were minor negative deviations 
observed just before and after the main positive deviation.

% Figure 1------------------------------------------------------
% --------------------------------------------------------------

% Table 1 ------------------------------------------------
\begin{table}[t]
%\footnotesize
%\small
%\centering
\caption{Factors used for error-bar normalisation.\label{table:one}}
%\begin{tabular}{lllllll}
\begin{tabular*}{\columnwidth}{@{\extracolsep{\fill}}lllll}
\hline\hline
\multicolumn{1}{c}{Dataset}                     &
\multicolumn{1}{c}{$k$}                          &
\multicolumn{1}{c}{$\sigma_{\rm min}$ (mag)}     \\
\hline
 KMTC03   &  0.954   &   0.020   \\
 KMTC43   &  0.959   &   0.020   \\
 KMTS03   &  0.995   &   0.020   \\
 KMTS43   &  0.964   &   0.020   \\
 KMTA03   &  0.922   &   0.030   \\
 KMTA43   &  0.938   &   0.050   \\
\hline                                                                                                                            
\end{tabular*}                             
\tablefoot{ ${\rm HJD}^\prime = {\rm HJD}- 2460000$.  }
\end{table}
% --------------------------------------------------------

Initially, the photometry data of the event were processed using the automated KMTNet pipeline,
which is based on the pySIS code developed by \citet{Albrow2009}. To ensure the data's optimal
quality for analysis, we performed additional photometry using the code developed by 
\citet{Yang2024}. In preparing the data for analysis, we normalised the error bars of the data 
to achieve consistency with the data scatter and to standardise the degree of freedom to unity 
for each dataset.  This error-bar normalisation procedure followed the protocol outlined in 
\citet{Yee2012}, that is,  $\sigma = k(\sigma_{\rm min} + \sigma_0)^{1/2}$, where $\sigma_{\rm min}$ 
is a factor used to consider the data scatter, and $k$ is the rescaling factor.  Table~\ref{table:one} 
list the values of $\sigma_{\rm min}$ and $k$ for the individual datasets.  For a subset of the 
KMTC data, we conducted additional photometry using the pyDIA code \citep{Albrow2017} for the 
source colour measurement. The detailed procedure for the source colour determination is described 
in Sect.~\ref{sec:four}.

\section{Lensing light curve analysis}\label{sec:three}

The pattern of the anomaly observed in the light curve of KMT-2024-BLG-1044 offers several 
crucial clues about its origin. Firstly, the rapid rise and fall of the source flux during 
the main feature of the anomaly suggest that it was likely caused by the source crossing a 
caustic created by a companion to the lens.  This interpretation is further supported by 
the brief negative deviations seen before and after the main feature.  Secondly, the absence 
of a characteristic U-shaped pattern between the two caustic spikes suggests that the 
source crossed a caustic tip, where the width of the tip is comparable to or narrower than 
the size of the source.  Lastly, the very brief duration of the anomaly implies that the lens 
companion is probably a very low-mass object.

% Figure 2------------------------------------------------------
\begin{figure}[t]
\includegraphics[width=\columnwidth]{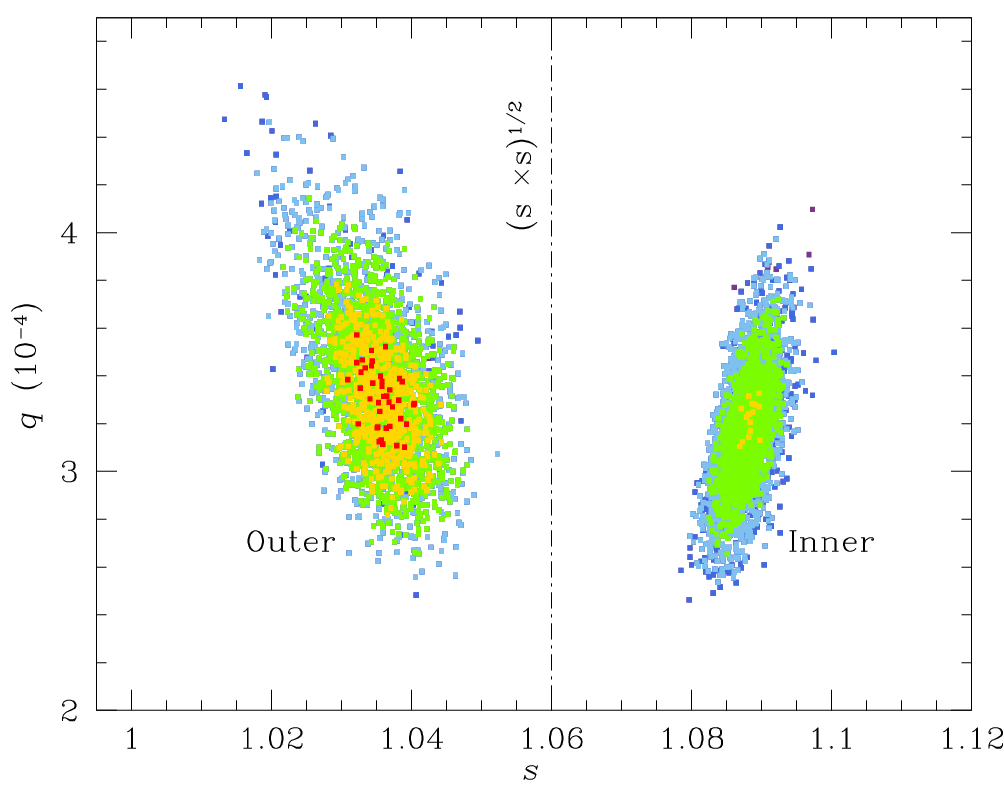}
\caption{
Scatter plots of points in the MCMC chain on the $(s, q)$ parameter plane.  The colour scheme 
is configured to represent points with $< 1\sigma$ (red), $< 2\sigma$ (yellow), $< 3\sigma$ 
(green), $< 4\sigma$ (cyan), and $< 5\sigma$ (blue). The dot-dashed vertical line in the left 
panel indicates the geometric mean of $s_{\rm in}$ and $s_{\rm out}$.
}
\label{fig:two}
\end{figure}
% --------------------------------------------------------------

Considering these characteristics of the anomaly, we conducted a binary-lens single-source (2L1S)
modelling of the event. The aim of this modelling was to determine the best-fit lensing parameters
(lensing solution) that characterise the lens system. In a 2L1S lensing event, the light curve 
is characterised by seven fundamental parameters. Among these, three parameters $(t_0, u_0, \te)$ 
depict the lens-source approach: $t_0$ denotes the time of the closest approach of the source to 
the lens, $u_0$ represents the lens-source separation at that time (impact parameter), and $\te$ 
is the event timescale. The timescale is defined as the duration required for the source to 
traverse the angular Einstein radius ($\thetae$), with the impact parameter $u_0$ scaled to 
$\thetae$. The binary lens system is described by two additional parameters $(s, q)$: $s$ 
represents the projected separation between the lens components ($M_1$ and $M_2$), and $q$ 
denotes the mass ratio between the lens components. The separation parameter $s$ is also scaled 
to $\thetae$. The parameter $\alpha$ represents the angle between the direction of the source 
motion and the binary-lens axis (source trajectory angle).  The final parameter, $\rho$, is 
defined as the ratio of the angular source radius ($\theta_*$) to the angular Einstein radius, 
that is, $\rho = \theta_*/\thetae$. This parameter characterises finite source magnifications 
during the caustic caustic crossings.

Finding a lensing solution for a 2L1S event using a downhill approach is very challenging due 
to the complexity of the $\chi^2$ surface in the parameter space. To overcome this difficulty, 
we employed a modelling strategy that combines grid and downhill methods. In this strategy, we 
searched for the binary-lens parameters $(s, q)$ using a grid approach. We set multiple starting 
values of $\alpha$, evenly distributed in the range of $0 < \alpha \leq 2\pi$, and find the other 
parameters via a downhill approach.  In the downhill approach, we used the Markov chain Monte Carlo 
(MCMC) algorithm with an adaptive step-size Gaussian sampler \citep{Doran2004}. In the second 
step, we constructed a $\chi^2$ map on the $s$--$q$ parameter plane to identify local solutions.  
In the third step, we refined each local solution by allowing all parameters to vary. This involves 
adjusting each parameter individually to ensure an optimal fit of each model to the data. In the 
final step, we compared the local solutions and selected a global solution for the event. If there 
is significant degeneracy among the local solutions, we present all of them. During the modelling, 
we computed finite magnifications using the map-making method outlined in \citet{Dong2006}.

% Figure 3------------------------------------------------------
\begin{figure}[t]
\includegraphics[width=\columnwidth]{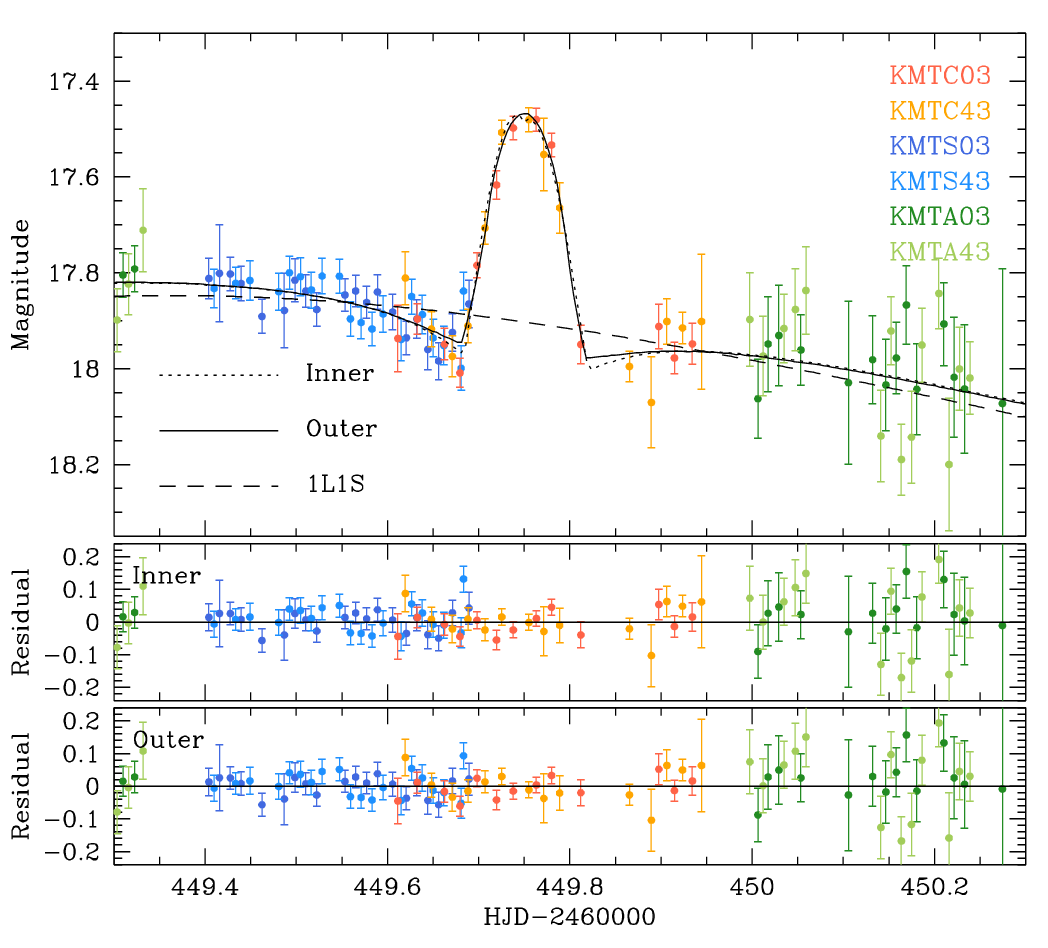}
\caption{
Model curves of the inner and outer 2L1S solutions and their residuals in the region
around the anomaly.
}
\label{fig:three}
\end{figure}
% --------------------------------------------------------------

% Figure 4------------------------------------------------------
\begin{figure}[t]
\includegraphics[width=\columnwidth]{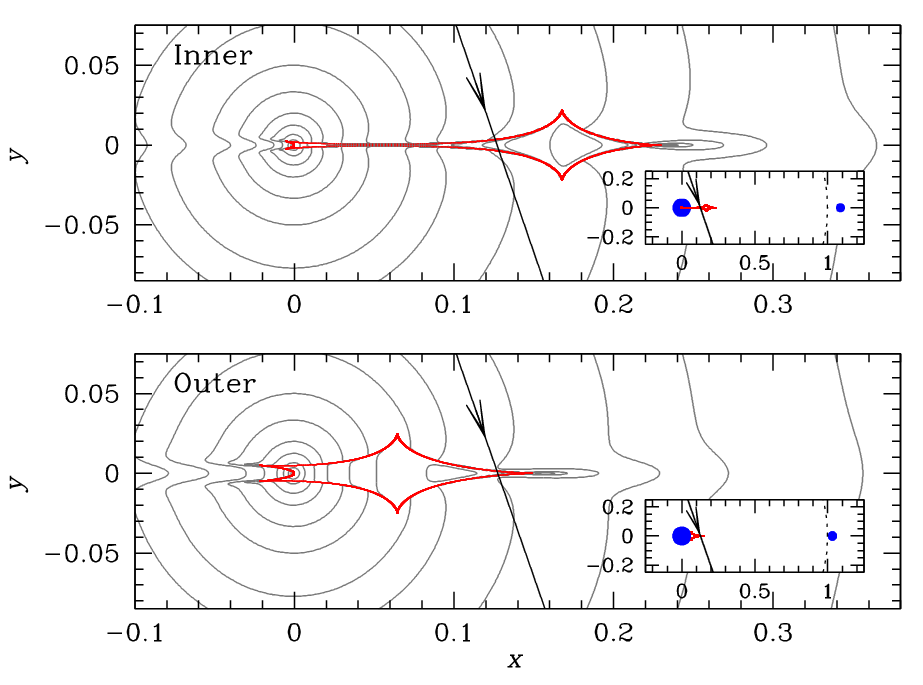}
\caption{
Configurations of the inner and outer solutions. In each panel, the red figure depicts the
caustic, and the arrowed line represents the source trajectory. The grey curves surrounding 
the caustic indicate the equi-magnification contours.  The inset provides a zoomed-out view 
showing the positions of the lens components, denoted by blue dots. The dotted curve 
in the inset illustrates the Einstein ring.
}
\label{fig:four}
\end{figure}
% --------------------------------------------------------------

% Table 2 ------------------------------------------------
\begin{table}[t]
%\footnotesize
%\small
%\centering
\caption{Best-fit lensing parameters.\label{table:two}}
%\begin{tabular}{lllllll}
\begin{tabular*}{\columnwidth}{@{\extracolsep{\fill}}lllll}
\hline\hline
\multicolumn{1}{c}{Parameter}         &
\multicolumn{1}{c}{Inner}             &
\multicolumn{1}{c}{Outer}             \\
\hline
 $\chi^2$                   &  $3780.0             $     &   $3776.6             $  \\
 $t_0$ (HJD$^\prime$)       &  $449.3714 \pm 0.0094$     &   $449.3686 \pm 0.0085$  \\
 $u_0$ (10$^{-2}$)          &  $0.1202   \pm 0.0053$     &   $0.1204   \pm 0.0052$  \\
 $\te$ (days)               &  $9.17     \pm 0.32  $     &   $9.13     \pm 0.29  $  \\
 $s$                        &  $1.0883   \pm 0.0027$     &   $1.0327   \pm 0.0054$  \\
 $q$ (10$^{-4}$)            &  $3.125    \pm 0.248 $     &   $3.350    \pm 0.316 $  \\
 $\alpha$ (rad)             &  $4.3822   \pm 0.0087$     &   $4.3771   \pm 0.0080$  \\
 $\rho$ (10$^{-3}$)         &  $3.46     \pm 0.30  $     &   $5.72     \pm 0.31  $  \\
\hline                                                                                                                            
\end{tabular*}                             
\tablefoot{ ${\rm HJD}^\prime = {\rm HJD}- 2460000$.  }
\end{table}
% --------------------------------------------------------

Detailed modelling of the light curve yielded a pair of degenerate solutions. The binary parameters 
of the solutions are $(s, q)_{\rm inner} \sim (1.09, 3.13\times 10^{-4})$ for one solution and 
$(s, q)_{\rm outer} \sim (1.03, 3.35\times 10^{-4})$ for the other.  The very low companion-to-primary 
mass ratio indicates that the companion is a planetary-mass object. Additionally, the proximity of 
the binary separation to unity suggests that the planet lies near the Einstein radius of the primary.  
We designate the individual solutions as `inner' and `outer' and explain the rationale for these 
terms in the following paragraph.  In Fig.~\ref{fig:two} we present a scatter plot of points 
in the MCMC chain on the $s$--$q$ parameter plane, showing that the two solutions are distinct 
despite the similarity in the $s$ and $q$ values.  In Table~\ref{table:two}, we list the full 
lensing parameters of the solutions along with the $\chi^2$ values of the fits. Among the 
determined lensing parameters, we highlight the relatively short timescale, which is less than 
10 days.  Because the event timescale is proportional to the square root of the lens mass, the 
short timescale of the event suggests that the lens mass is likely small.  In Fig.~\ref{fig:three} 
we present the model curves of the two solutions in the region around the anomaly.  It was found that 
the outer solution is favoured over the inner solution, but the difference in $\chi^2$ between the 
fits of the two solutions, $\Delta\chi^2 = 3.6$, is minor.

\citet{Gaudi1998} noted that a short-term positive anomaly could be caused by the approach of 
a faint companion to the source.  We explored this interpretation by conducting an additional 
modelling under single-lens and binary-source (1L2S) configuration.  Our findings indicate that 
the 1L2S interpretation is ruled out with a $\Delta\chi^2$ value of 1248.0.

Figure~\ref{fig:four} illustrates the lens-system configurations for both the inner and outer 
2L1S solutions. In each configuration, the caustics exhibit a resonant form, with the central 
and planetary caustics merging into a single closed curve. For the inner solution, the source 
passed through the inner side of the planetary caustic, while for the outer solution, it traversed 
the outer side. Therefore, we refer to these configurations as `inner' and `outer'. This degeneracy 
was first pointed out by \citet{Gaudi1997} to highlight the similarities in anomalies caused by 
source passages over the inner and outer sides of well-detached planetary caustics. \citet{Yee2021}, 
\citet{Zhang2022a}, and \citet{Zhang2022b} further demonstrated that this degeneracy also applies 
to planetary signals resulting from semi-detached resonant caustics.

It is known that the binary separations of the pair of solutions under the inner--outer degeneracy
($s_{\rm in}$ and $s_{\rm out}$) follow the relation
\begin{equation}
s_\pm^\dagger = \sqrt{s_{\rm in} \times s_{\rm out}} = 
{1 \over 2}
\left(\sqrt{u_{\rm anom}^2+4}\pm u_{\rm anom}\right).
\label{eq1}
\end{equation}
Here $u^2_{\rm anom} = \tau^2_{\rm anom} + u_0^2$, $\tau_{\rm anom}=(t_{\rm anom}-t_0)/\te$, 
$t_{\rm anom}$ represents the time of the anomaly, and the `+' and `$-$' signs in the first 
and later terms are for anomalies exhibiting a bump feature (positive anomaly) and a dip 
feature (negative anomaly), respectively \citep{Hwang2022, Gould2022}.  For KMT-2024-BLG-1044, 
the anomaly exhibits a bump feature, and thus the sign is `+'.  The lensing parameters $(t_0, 
u_0, \te, t_{\rm anom})=(449.37, 0.120, 9.15, 449.75)$ yield $s^\dagger =1.065$, which matches 
the geometric mean of $(s_{\rm in} \times s_{\rm out})^{1/2} =1.060$ well.  This confirms that 
the similarity between the two model curves originates from the inner--outer degeneracy.

% Figure 5------------------------------------------------------
\begin{figure}[t]
\includegraphics[width=\columnwidth]{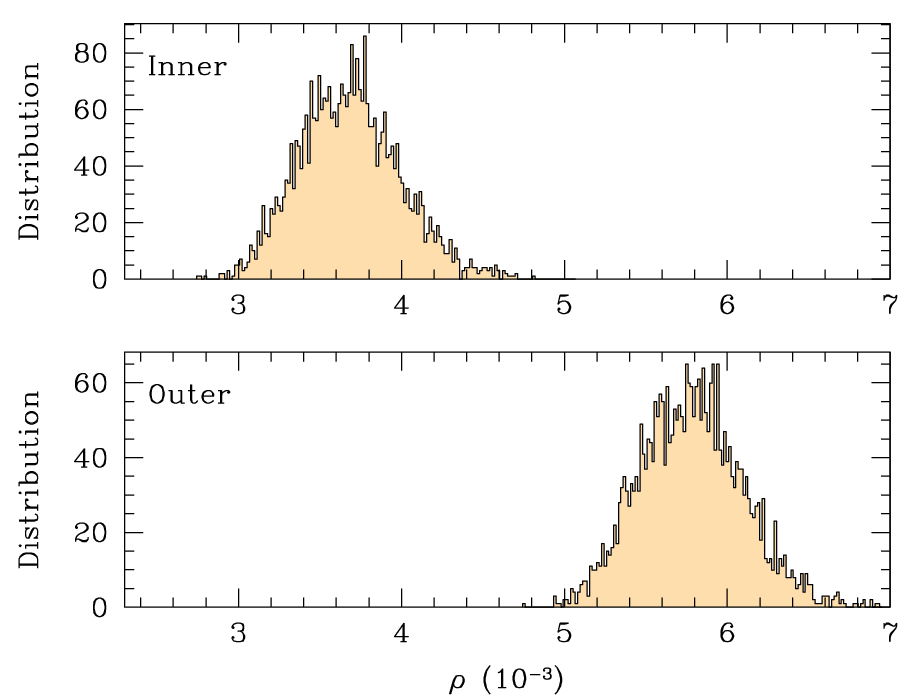}
\caption{
Distributions of the normalised source radius for the inner and outer solutions.
}
\label{fig:five}
\end{figure}
% --------------------------------------------------------------

\begin{figure}[t]
\includegraphics[width=\columnwidth]{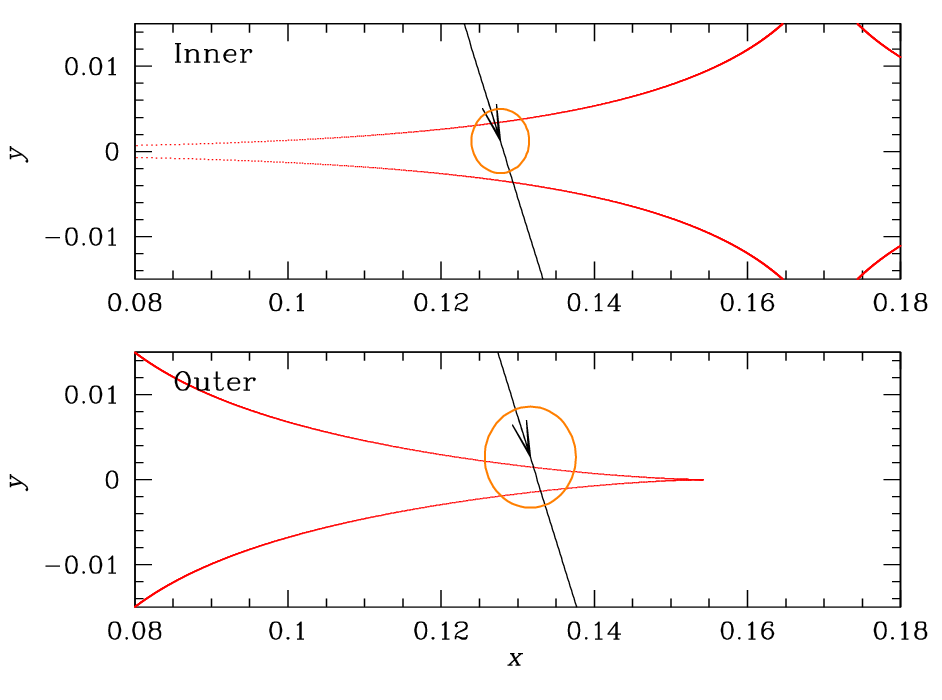}
\caption{
Zoomed-in view of the caustic crossings for the inner and outer solutions.  In each panel, the orange circle 
represents the source size scaled to the caustic size. 
}
\label{fig:six}
\end{figure}

Apart from the difference in their projected separations, there are substantial differences in 
the estimated values of the normalised source radii between the inner and outer solutions: 
$\rho_{\rm in}=(3.46\pm 0.30)\times 10^{-3}$ for the inner solution and $\rho_{\rm out} = (5.72
\pm 0.31)\times 10^{-3}$, for the outer solution. Figure~\ref{fig:five} shows the distributions 
for the MCMC points of the individual local solutions.  In Fig.~\ref{fig:six} we present 
the zoom of the caustic crossings for each solution, with the source size shown to scale.  The 
two solutions are caused by the different separations between the two caustic walls.  For the 
inner solution, the two walls are farther apart, and thus the empirically determined width of 
the bump can be caused by a smaller source size that is kept relatively highly magnified for a 
long time by the two caustics.  But for the outer solution, the two walls are closer together, 
and so the source size must be bigger to keep the same size bump.  The angular Einstein radius 
$\thetae$ is derived from the measured normalised source radius by the relation
\begin{equation}
\thetae = {\theta_* \over \rho},
\label{eq2}
\end{equation}
where $\theta_*$ represents the angular source radius. As a result, the difference in the value 
of $\rho$ between the two solutions leads to different values of $\thetae$.  Because the angular 
Einstein radius is proportional to the square root of the lens mass, the lens masses expected 
from the two solutions will be different.  We provide a detailed discussion on the estimation of 
the angular Einstein radius and lens mass in Sects.~\ref{sec:four} and \ref{sec:five}, respectively.

% Figure 6------------------------------------------------------
% --------------------------------------------------------------

\section{Source star and angular Einstein radius}\label{sec:four}

In this section we specify the source star of the event. Specifying the source is important not only
for providing a comprehensive overview of the event but also for estimating the angular Einstein
radius.

We specified the source star of the event using the methodology outlined in \citet{Yoo2004}. First, 
we estimated the instrumental source magnitudes in the $I$ and $V$ passbands by linearly regressing 
the datasets, processed with the pyDIA photometry code \citep{Albrow2017}, against the model light 
curve.  Next, we placed the source in the colour-magnitude diagram (CMD) of stars near the source, 
constructed using the same pyDIA code. Finally, we calibrated the $I$- and $V$-band source magnitudes 
using the centroid of the red giant clump (RGC) in the CMD. The RGC centroid can be used for calibration 
because its $(V-I)$ colour and $I$-band magnitude are known from previous studies conducted by 
\citet{Bensby2013} and \citet{Nataf2013}.

% Figure 7------------------------------------------------------
\begin{figure}[t]
\includegraphics[width=\columnwidth]{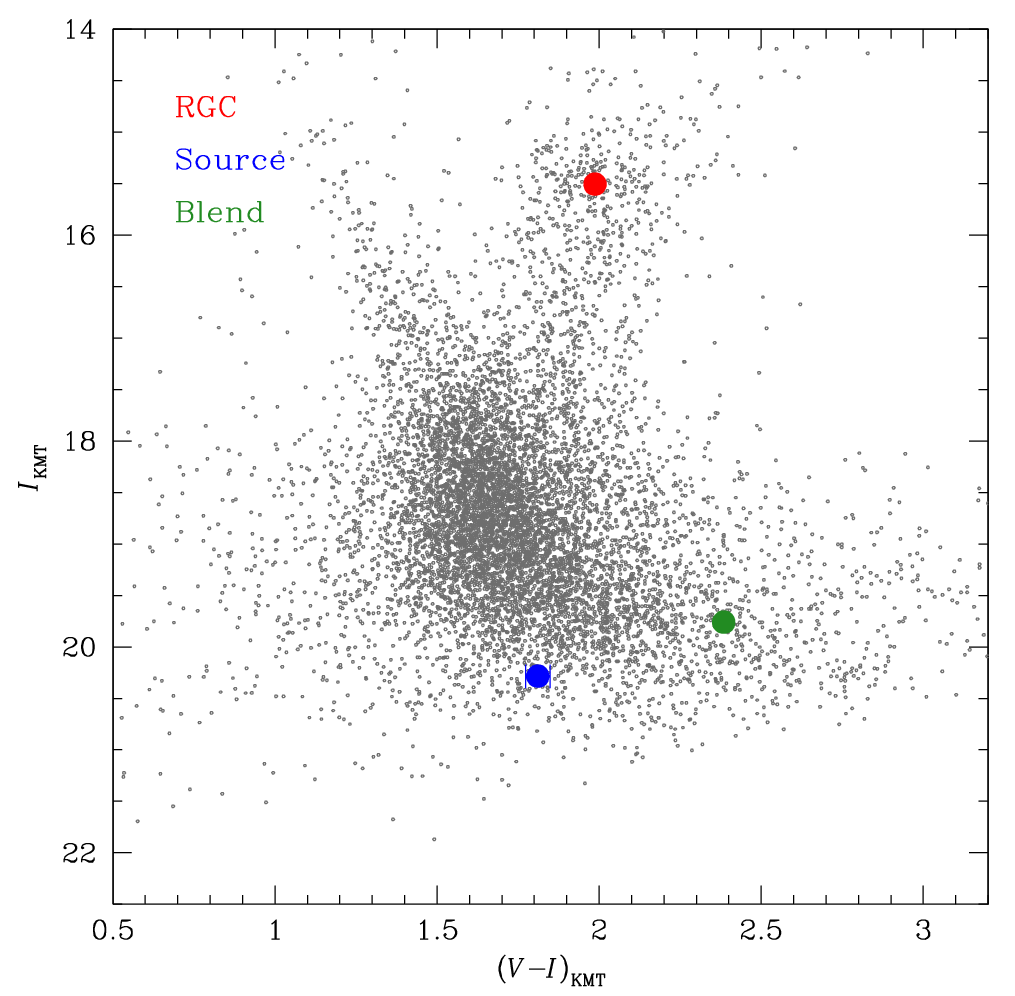}
\caption{
Locations of the source and the centroid of the RGC in the instrumental CMD of stars in the vicinity of the source. Also marked is the location of the blend.
}
\label{fig:seven}
\end{figure}
% --------------------------------------------------------------

Figure~\ref{fig:seven} shows the location of the source in the instrumental CMD of stars lying 
near the source.  The measured instrumental colour and magnitude of the source are
\begin{equation}
(V-I, I)_{\rm S} 
= (1.811 \pm 0.038, 20.285 \pm 0.006).
\label{eq3}
\end{equation}
By measuring the offset, $\Delta(V-I, I) = (V-I, I)_{\rm S} - (V-I, I)_{\rm RGC}$, between the 
source and RGC centroid, lying at $(V-I, I)_{\rm RGC} = (1.987, 15.506)$, the de-reddened source 
colour and magnitude are estimated as
\begin{equation}
\eqalign{
(V-I, I)_{{\rm S},0}  & = (V-I, I)_{{\rm RGC},0} + \Delta(V-I, I) \cr
                      & = (0.884 \pm 0.055, 19.118 \pm  0.021).
}
\label{eq4}
\end{equation}
Here $(V-I, I)_{{\rm RGC},0} = (1.060, 14.339)$ represent the de-reddened colour and magnitude of 
the RGC centroid.  The estimated colour and magnitude indicate that the source is an early K-type 
main-sequence star. In the CMD, we also mark the location of the blend with a green dot.
It is important to note that the instrumental colour and magnitudes are not precisely scaled, 
as the calculation of $(V-I, I)_{{\rm S},0}$ relies only on the offset $\Delta (V-I, I)$, 
not on the absolute calibration of $(V-I, I)$ and $(V-I, I)_{\rm RGC}$.

Based on the measured source colour and magnitude, we then estimated the angular radius of the 
source.  For this, we first converted $V-I$ colour into $V-K$ colour using the \citet{Bessell1988} colour-colour relation and then interpolated the angular source radius from the \citet{Kervella2004} 
relation between $V-K$ and $\theta_*$. This yields the angular source radius of
\begin{equation}
\theta_* = (0.573 \pm 0.051)~\mu{\rm as}.
\label{eq5}
\end{equation}
With the estimated source radius, the angular radius of the Einstein ring was estimated using 
the relation in Eq.~(\ref{eq2}) as
\begin{equation}
\thetae = 
\begin{cases}
(0.166 \pm 0.021)~{\rm mas},  & \text{(inner solution)},        \\
(0.100 \pm 0.010)~{\rm mas},  & \text{(outer solution)}. \\
\end{cases}
\label{eq6}
\end{equation}
It is important to note that the inner and outer solutions yield different values of the normalised 
source radius, which in turn result in different values of angular Einstein radius. With the 
determined value of $\thetae$ together with the event timescale, estimated from the light curve 
modelling, the relative proper motion between the lens and source is determined as
\begin{equation}
\mu = {\thetae \over \te} = 
\begin{cases}
(6.59 \pm 0.82)~{\rm mas/yr},  & \text{(inner solution)}, \\
(4.01 \pm 0.42)~{\rm mas/yr},  & \text{(outer solution)}. \\
\end{cases}
\label{eq7}
\end{equation}

% Figure 8------------------------------------------------------
\begin{figure}[t]
\includegraphics[width=\columnwidth]{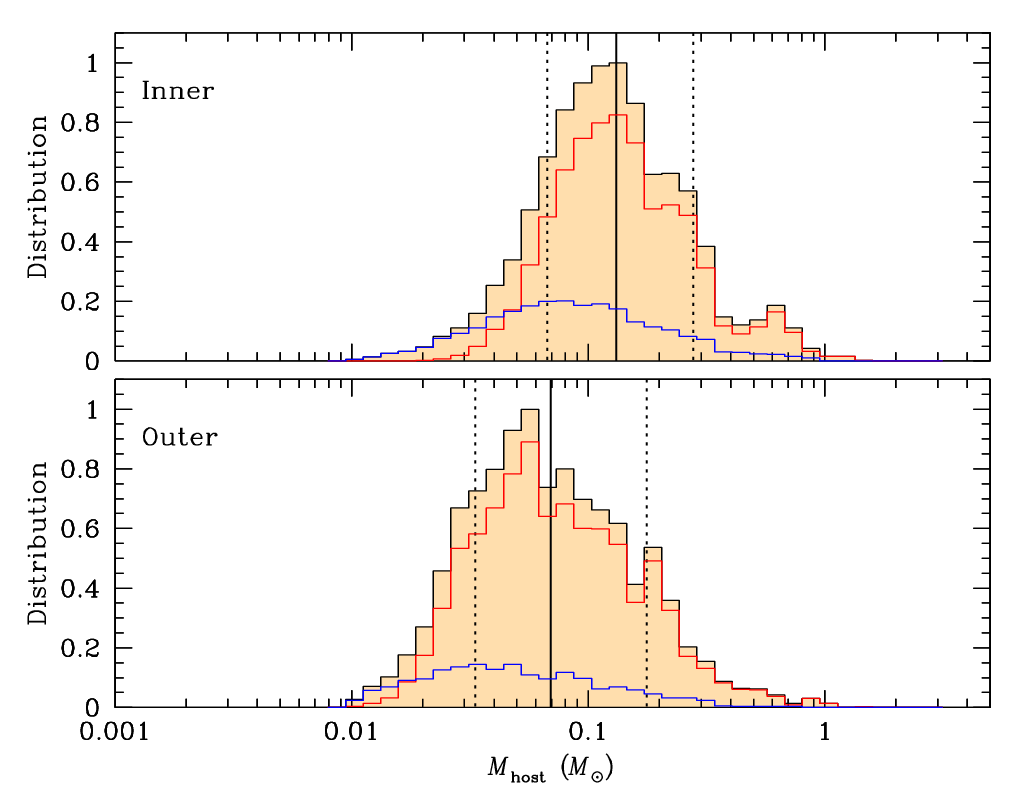}
\caption{
Bayesian posteriors for the host mass of the planetary system KMT-2024-BLG-1044L. The distributions 
shown in the upper and lower panels are based on the inner and outer solutions, respectively. In 
each panel, the vertical solid line indicates the median value of the distribution, while the two 
dotted vertical lines denote the 1$\sigma$ range. The distributions, depicted in blue and red, 
represent the contributions from the disc and bulge lens populations, respectively. The black 
curve represents the sum of these two contributions.
}
\label{fig:eight}
\end{figure}
% --------------------------------------------------------------

\section{Physical lens parameters}\label{sec:five}

The mass ($M$) and distance ($\dl$) of a microlens are constrained by measuring three lensing 
observables: $\te$, $\thetae$, and $\pie$. Here the last observable $\pie$ denotes the microlens 
parallax. These observables are related to $M$ and $\dl$ by the relations
\begin{equation}
\te = {\thetae \over \mu}; \qquad
\thetae = \sqrt{\kappa M \pi_{\rm rel}}; \qquad
\pie = \left( {\pi_{\rm rel}\over \thetae}\right) \left( {\muvec \over \mu} \right),
\label{eq8}
\end{equation}
where $\kappa = 4G/(c^2{\rm AU})$, $\pi_{\rm rel} = \pi_{\rm L} - \pi_{\rm S} = {\rm AU}(1/\dl 
-  1/\ds)$ is the relative lens-source parallax, $\ds$ represents the distance to the source, 
and $\muvec$ denotes the vector of the relative lens-source proper motion \citep{Gould2000}.  
For KMT-2024-BLG-1044, the event timescale and the angular Einstein radius were securely measured, 
but the microlens parallax could not be determined due to the combination of relatively large 
photometric uncertainties in the data and the short duration of the event. Consequently, we 
estimated the physical parameters of the lens through Bayesian analysis, using priors on the mass, 
location, and motion of lens objects within the Galaxy, and incorporating constraints from the 
measured observables.

In the Bayesian analysis, we first generated a large number of artificial events through a
Monte Carlo simulation. In this simulation, the lens mass was derived from a mass-function
model, while the distances to the lens and source, as well as the transverse lens-source
speed ($v_\perp$), were derived from a Galaxy model.  We used the mass function model from 
\citet{Jung2022} and the Galaxy model from \citet{Jung2021}.  In the Galaxy model, the bulge 
density profile is represented by a triaxial distribution, while the disc matter density is 
described by a double-exponential distribution. For the stellar and BD mass functions, 
the initial mass function from \citet{Chabrier2003} was applied to the bulge lens population, 
whereas the present-day mass function from the same source was used for the disc lens population.  
With the assigned physical parameters $(M_i, D_{{\rm L},i}, D_{{\rm S},i}, v_{\perp,i})$ for 
each artificial event, we computed the corresponding lensing observables using the relations 
$t_{{\rm E},i}= D_{{\rm L},i} \theta_{{\rm E},i} /v_{\perp,i}$, $\theta_{{\rm E},i} =  (\kappa 
M_i \pi_{{\rm rel},i})^{1/2}$, and $\pi_{{\rm rel},i} = {\rm AU}(1/D_{{\rm L},i}-1/D_{{\rm S},i})$.  
In the second step, we constructed posterior distributions of the lens mass and distance by 
imposing a weight to the event of
\begin{equation}
w_i = \exp \left(  {\chi_i^2 \over 2}\right); \qquad
\chi_i^2  = 
        { (t_{{\rm E},i}-\te )^2 \over \sigma^2(\te)} +
        { (\theta_{{\rm E},i}-\thetae )^2 \over \sigma^2(\thetae)}.
\label{eq9}
\end{equation}
Here $(\te, \thetae)$ represent the measured values of the lensing observables, and $[\sigma(\te), 
\sigma(\thetae)]$ are their measurement uncertainties.  In the Bayesian analysis, we imposed an 
additional constraint, that the lens flux be less than the blended flux.  However, this constraint 
had little effect on the posteriors because the estimated lens mass is very low.

% Figure 9 -----------------------------------------------------
\begin{figure}[t]
\includegraphics[width=\columnwidth]{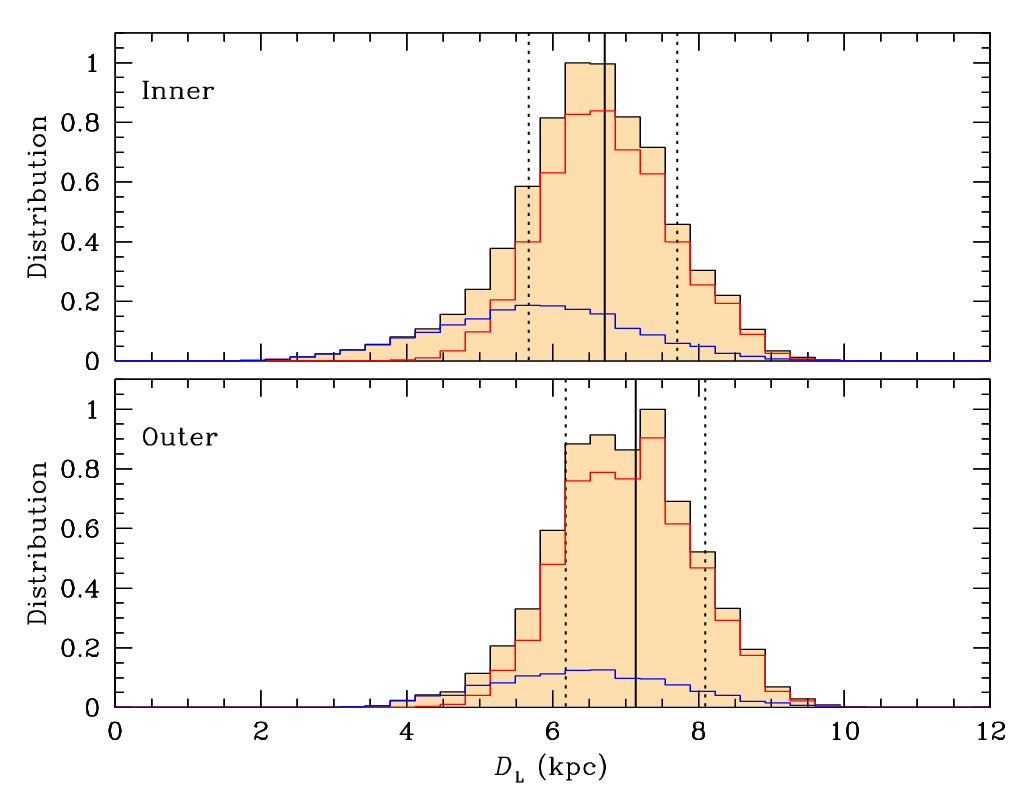}
\caption{
Bayesian posteriors for the distance to the lens. The notations are same in 
Fig.~\ref{fig:eight}.
}
\label{fig:nine}
\end{figure}
% --------------------------------------------------------------

% Table 3 ------------------------------------------------
\begin{table*}[t]
%\footnotesize
\small
%\centering
\caption{Candidate microlensing planetary systems with BD hosts.\label{table:three}}
\begin{tabular}{llllllllll}
%\begin{tabular}{\columnwidth}{@{\extracolsep{\fill}}lllcc}
\hline\hline
\multicolumn{1}{c}{Event}           &
\multicolumn{1}{c}{$q$}             &
\multicolumn{1}{c}{$\te$ (days)}    &
\multicolumn{1}{c}{$\thetae$ (mas)} &
\multicolumn{1}{c}{$\pie$}          &
\multicolumn{1}{c}{$M_{\rm h}$}     &
\multicolumn{1}{c}{$M_{\rm p}$}     &
\multicolumn{1}{c}{Reference}      \\
\hline
MOA-2007-BLG-192      &  0.0002    &  70       &  --         &  1.5              &  0.060 $M_\odot$  &   3.3   $M_{\rm E}$      &  \citet{Bennett2008}     \\ 
MOA-2011-BLG-262      &  0.0005    &  3.87     &  0.136      &  --               &  3.2  $M_{\rm J}$ &   10.5  $M_{\rm E}$      &  \citet{Bennett2014}     \\ 
OGLE-2012-BLG-0358    &  0.12      &  25.64    &  0.29       &  1.5              &  0.024 $M_\odot$  &   1.82  $M_{\rm J}$      &  \citet{Han2013}         \\ 
MOA-2013-BLG-605      &  0.0035    &  20       &  0.48       &  1.5--3.7         &  0.15  $M_\odot$  &   18    $M_{\rm E}$      &  \citet{Sumi2016}        \\ 
                      &            &           &             &  --               &  0.05  $M_\odot$  &   6     $M_{\rm E}$      &                          \\
OGLE-2015-BLG-1771    &  0.00538   &  4.28     &  0.111      &  $\sim$0.4 -- 0.5 &  0.077 $M_\odot$  &   0.43  $M_{\rm J}$      &  \citet{Zhang2020}       \\ 
                      &            &           &  0.132      &  --               &  0.086 $M_\odot$  &   0.40  $M_{\rm J}$      &                          \\ 
                      &            &           &  0.079      &  --               &  0.055 $M_\odot$  &   2.63  $M_{\rm J}$      &                          \\ 
OGLE-2016-BLG-1195    &  0.000055  &  10.2     &  0.28       &  --               &  0.078 $M_\odot$  &   1.43  $M_{\rm E}$      &  \citet{Shvartzvald2017} \\
KMT-2016-BLG-1820     &  0.11300   &  4.81,    &  0.123      &  --               &  0.039 $M_\odot$  &   4.57  $M_{\rm J}$      &  \citet{Jung2018}        \\ 
KMT-2016-BLG-2142     &  0.21      &  5.2/6.1  &  0.122      &  --               &  0.073 $M_\odot$  &   15.49 $M_{\rm J}$      &  \citet{Jung2018}        \\ 
KMT-2016-BLG-2605     &  0.0120    &  3.41     &  0.116      &  --               &  0.064 $M_\odot$  &   0.771 $M_{\rm J}$      &  \citet{Ryu2021b}         \\ 
OGLE-2017-BLG-1522    &  0.016     &  7.53     &  0.065      &  --               &  0.045 $M_\odot$  &   0.75  $M_{\rm J}$      &  \citet{Jung2018}        \\
KMT-2018-BLG-0748     &  0.0120    &  4.38     &  0.111      &  --               &  0.087 $M_\odot$  &   0.19  $M_{\rm J}$      &  \citet{Han2020}          \\ 
OGLE-2018-BLG-0677    &  0.000079  &  4.94     &  > 0.049    &  --               &  0.12 $M_\odot$   &   3.96  $M_{\rm E}$      &  \citet{Herrera2020}     \\
KMT-2021-BLG-0371     &  0.079     &  6.53     &  0.135      &  --               &  0.09  $M_\odot$  &   7.70  $M_{\rm J}$      &  \citet{Kim2021b}         \\ 
KMT-2021-BLG-1554     &  0.0014    &  5.1      &  0.10       &  --               &  0.08  $M_\odot$  &   0.12  $M_{\rm J}$      &  \citet{Han2022}         \\ 
KMT-2024-BLG-1044     &  0.00031   &  9.1      &  0.166      &  --               &  0.13  $M_\odot$  &   13.7  $M_{\rm E}$      &  This work               \\ 
                      &  0.00034   &           &  0.100      &  --               &  0.07  $M_\odot$  &   7.8   $M_{\rm E}$      &                          \\ 
\hline                                                                                                                              
\end{tabular}                             
%\tablefoot{ ${\rm HJD}^\prime = {\rm HJD}- 2460000$.  }
\end{table*}
% --------------------------------------------------------

In Figs.~\ref{fig:eight} and \ref{fig:nine} we present the posterior distributions of the 
host mass and distance to the planetary system KMT-2024-BLG-1044L, respectively.  We present 
two sets of distributions derived from separate analyses based on the inner and outer solutions, 
as these solutions result in substantially different values of $\thetae$. The estimated masses 
of the host and planet are \begin{equation}
M_{\rm h} = 
\begin{cases}
0.131^{+0.146}_{-0.064}~M_\odot,  & \text{(inner solution),} \\[0.9ex]
0.069^{+0.080}_{-0.036}~M_\odot,  & \text{(outer solution),} \\
\end{cases}
\label{eq10}
\end{equation}
and 
\begin{equation}
M_{\rm p} = 
\begin{cases}
13.68^{+15.23}_{-6.71}~M_{\rm E},  & \text{(inner solution)}, \\[0.9ex]
7.75^{+12.03}_{-4.04} ~M_{\rm E},  & \text{(outer solution)}, \\
\end{cases}
\label{eq11}
\end{equation}
respectively.  Here we set the lower and upper limits of the physical parameters as the 16th and 84th 
percentiles of the Bayesian posteriors.  The estimated physical parameters suggest that the planetary 
system consists of a host near the boundary between a star and a BD, with a planet possessing 
a mass smaller than that of Uranus.  The estimated distance to the planetary system 
is
\color{black}
\begin{equation}
\dl = 
\begin{cases}
6.71^{+0.99}_{-1.04}~{\rm kpc},  & \text{(inner solution)}, \\[0.9ex]
7.14^{+0.95}_{-0.96}~{\rm kpc},  & \text{(outer solution)}. \\
\end{cases}
\label{eq12}
\end{equation}
For the inner solution, the relative probabilities suggest a 25\% chance of the lens lying in the 
disc and a 75\% chance in the bulge. For the outer solution, these probabilities indicate a 17\% 
chance in the disc versus 83\% in the bulge. These probabilities indicate that the lens is more 
likely to be located in the bulge.  The contributions from the disc and bulge lens populations were 
determined by summing the event rates of individual simulated microlensing events produced by lenses 
from each population.

\section{Summary and discussion}\label{sec:six}

We analysed microlensing data to understand the nature of the very short-term anomaly 
that appeared near the peak of the short-timescale microlensing event KMT-2024-BLG-1044. 
Detailed modelling of the light curve confirmed the planetary origin of the anomaly and revealed 
two possible solutions, due an to inner--outer degeneracy. The measured planet-to-host 
mass ratio is $3.1 \times 10^{4}$ according to the inner solution and $3.4 \times 10{^4}$ according to the outer 
solution.  Using Bayesian analysis, constrained by both the short event timescale and the small 
angular Einstein radius, we have determined that the lens system is a planetary system. The mass 
estimates derived from this analysis suggest two possible interpretations for the nature of the 
host. According to the inner solution, the host star is likely a low-mass stellar object. However, 
the outer solution suggests that the host is a BD, with its mass falling into the 
substellar range. Resolving this degeneracy between the two possibilities presents significant 
challenges, primarily due to the limitations of the photometric data combined with the absence 
of spectroscopic data.

Moreover, the difficulty is compounded by the current lack of comprehensive knowledge regarding 
planet formation mechanisms in such systems, particularly for planets orbiting very low-mass 
stars and BDs. \citet{Matsuo2007} pointed out that additional information, such as the 
metallicity of the host star, could provide valuable constraints on the formation mechanism.  
Specifically, the two dominant models for planet formation, the core accretion model and the 
disc instability model, could be better differentiated with this kind of supplementary data.  
In the case of the KMT-2024-BLG-1044L system, however, it remains difficult to draw a definitive 
conclusion about the planet formation mechanism. This is because, while we have estimates for 
the masses of both the planet and its host, there is a lack of the critical additional information 
-- such as host star metallicity -- that would enable us to favour one formation mechanism over the 
other. Consequently, our current understanding of the system is limited, and further data would be 
required to make a more robust determination.

% Figure 10------------------------------------------------------
\begin{figure}[t]
\includegraphics[width=\columnwidth]{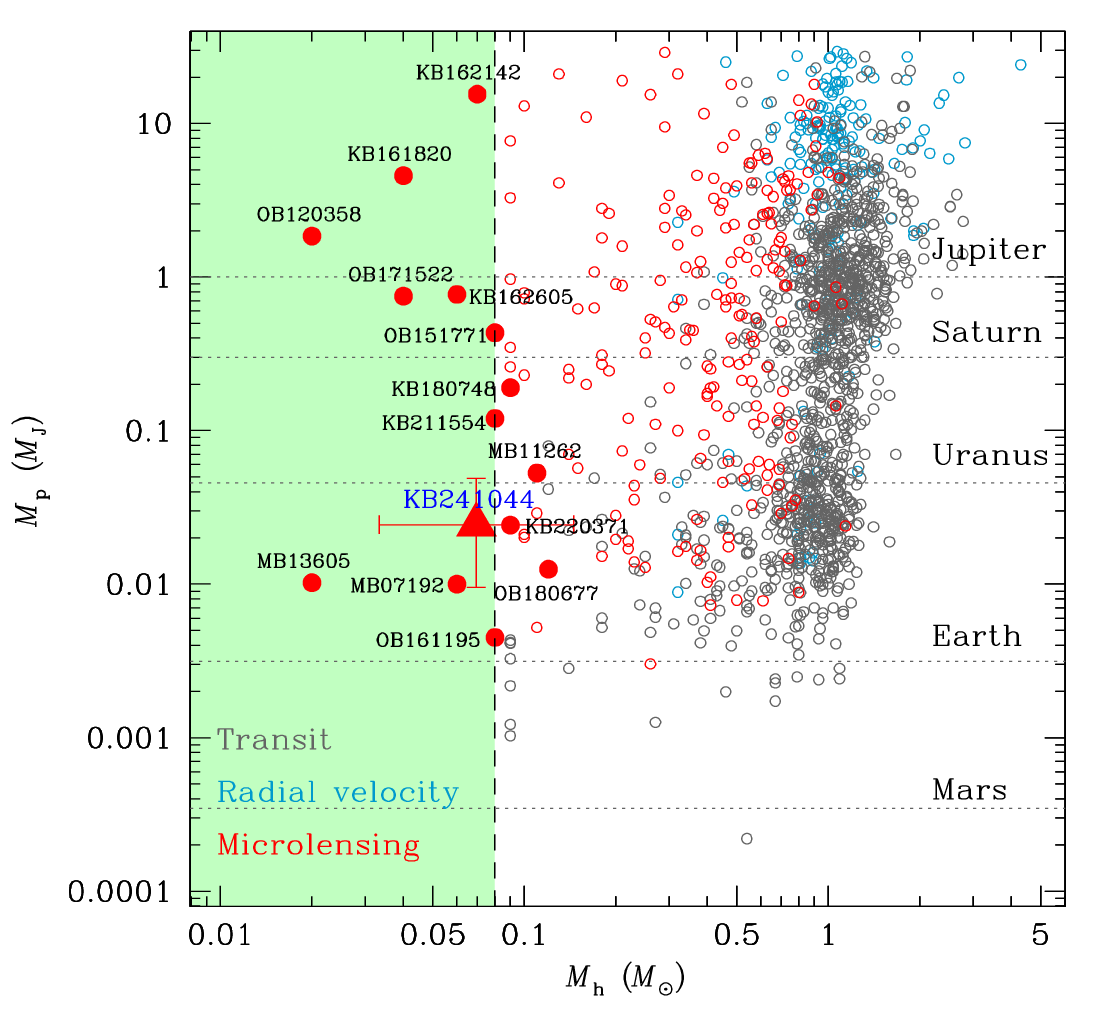}
\caption{
Distribution of extrasolar planets detected using the three major methods -- transit, radial 
velocity, and microlensing -- in the plane of the host and planet masses.  The green shade 
indicates the region where the host has a substellar mass.  Red-filled dots represent planetary 
systems with masses near or below the star-brown dwarf mass threshold.  The position of 
KMT-2024-BLG-1044L is marked with a triangle.
}
\label{fig:ten}
\end{figure}
% --------------------------------------------------------------

The discovery of the planetary system KMT-2024-BLG-1044L underscores the crucial role played 
by the microlensing technique in detecting planets that orbit substellar BDs or very low-mass 
stars. This is illustrated in Fig.~\ref{fig:ten}, which depicts the distribution of extrasolar 
planets detected using the three major methods: transit, radial velocity, and microlensing. 
The data for the planets in the plot -- 1149 transit planets, 195 radial velocity planets, and 
221 microlensing planets with measured masses -- were sourced from the NASA Exoplanet Archive.
Within this dataset, there are  15 planetary systems for which the host has a substellar mass, 
defined as $M_{\rm h} \lesssim 0.8~M_\odot$, or lying around the star--BD boundary.  To 
investigate the common characteristics of BD-host planets, we present Table~\ref{table:three}, 
which lists the planet and host masses, planet-to-host mass ratios, and lensing observables 
($\te$, $\thetae$, and $\pie$) for these planetary systems.  For events with multiple degenerate 
solutions, we list the values of $M_{\rm p}$ and $M_{\rm h}$ that correspond to the individual 
solutions.  Upon inspecting the lensing observables, we see that these planets can be broadly 
divided into two groups depending on how the BD nature of the host is identified.  The first 
group, which includes MOA-2007-BLG-192, OGLE-2012-BLG-0358, and OGLE-2016-BLG-1195, was 
identified through the measurement of the large microlens parallax. The second group, which 
includes OGLE-2015-BLG-1771, KMT-2016-BLG-1820, KMT-2016-BLG-2142, KMT-2016-BLG-2605, 
OGLE-2017-BLG-1522, KMT-2018-BLG-0748, OGLE-2018-BLG-0677, KMT-2021-BLG-0371, KMT-2021-BLG-1554, 
and KMT-2024-BLG-1044, was identified primarily through the combination of short event timescales 
and the small angular radii of the Einstein rings.  Notably, the discovery of all these planets 
has been made feasible through the unique capabilities of the microlensing method, particularly 
in detecting dark or faint celestial bodies.  This highlights the method's effectiveness in 
revealing planetary systems that might otherwise be challenging to detect using other observational 
methods.

\begin{acknowledgements}
%Work by C.H. was supported by the grants of National Research Foundation of Korea (2019R1A2C2085965). 
% Yee
J.C.Y. and I.-G.S. acknowledge support from U.S. NSF Grant No. AST-2108414. 
% Yossi Shvartzvald 
Y.S. acknowledges support from BSF Grant No. 2020740.
% KMTNet
This research has made use of the KMTNet system operated by the Korea Astronomy and Space
Science Institute (KASI) at three host sites of CTIO in Chile, SAAO in South Africa, and SSO in
Australia. Data transfer from the host site to KASI was supported by the Korea Research
Environment Open NETwork (KREONET). This research was supported by KASI under the R\&D
program (Project No. 2024-1-832-01) supervised by the Ministry of Science and ICT.
% Chinese researcher 
W.Z. and H.Y. acknowledge support by the National Natural Science Foundation of China (Grant
No. 12133005).
W. Zang acknowledges the support from the Harvard-Smithsonian Center for Astrophysics through
the CfA Fellowship. 
% OGLE 
%The OGLE has received funding from the National Science Centre, Poland, grant MAESTRO
%2014/14/A/ST9/00121 to AU.
% MOA
%The MOA project is supported by JSPS KAKENHI Grant Number JP24253004, JP26247023,JP16H06287 and JP22H00153.
\end{acknowledgements}


\begin{thebibliography}{}
% -------------
\bibitem[Albrow et al.(1998)]{Albrow1998} Albrow, M., Beaulieu, J.-P., Birch, P., et al. 1998, \apj, 509, 687
\bibitem[Albrow et al.(2009)]{Albrow2009} Albrow, M., Horne, K., Bramich, D.~M., et al.\ 2009, \mnras, 397, 2099
\bibitem[Albrow(2017)]{Albrow2017} Albrow, M.\ 2017, MichaelDAlbrow/pyDIA: Initial Release on Github,Versionv1.0.0, Zenodo, doi:10.5281/zenodo.268049
\bibitem[Alcock et al.(1993)]{Alcock1993} Alcock, C., Akerlof, C. W., Allsman, R. A., et al. 1993, Nature, 365, 621
\bibitem[Alcock et al.(1997)]{Alcock1997} Alcock, C., Allen, W. H., Allsman, R. A., et al. 1997, \apj, 491, 436
\bibitem[Bennett et al.(2008)]{Bennett2008} Bennett, D. P., Bond, I. A., Udalski, A., et al. 2008, \apj, 684, 663
\bibitem[Bennett et al.(2014)]{Bennett2014} Bennett, D. P., Batista, V., Bond, I. A., et al. 2014, \apj, 785, 155
\bibitem[Bensby et al.(2013)]{Bensby2013} Bensby, T., Yee, J.~C., Feltzing, S., et al.\ 2013, \aap, 549, A147
\bibitem[Bessell \& Brett(1988)]{Bessell1988} Bessell, M. S., \& Brett, J. M. 1988, \pasp, 100, 1134
\bibitem[Bond, et al.(2001)]{Bond2001} Bond, I. A., Abe, F., Dodd, R. J., et al. 2001, \mnras, 327, 868
\bibitem[Bond et al.(2004)]{Bond2004} Bond, I. A., Udalski, A., Jaroszy\'nski, M., et al. 2004, \apj, 606, L155
\bibitem[Chabrier(2003)]{Chabrier2003} Chabrier, G. 2003, \apjl, 586, L133
\bibitem[Dong et al.(2006)]{Dong2006} Dong, S., DePoy, D. L., Gaudi, B. S., et al. 2006, \apj, 642, 842
\bibitem[Doran \& Mueller(2004)]{Doran2004} Doran, M., \& Mueller, C. M. 2004, J. Cosmology Astropart. Phys., 09, 003
\bibitem[Gaudi(1998)]{Gaudi1998} Gaudi, B. S. 1998, \apj, 506, 533
\bibitem[Gaudi(2012)]{Gaudi2012} Gaudi, B. S. 2012, \araa, 50, 411
\bibitem[Gaudi  \& Gould(1997)]{Gaudi1997} Gaudi, B. S., \& Gould, A. 1997, \apj, 486, 85
\bibitem[Gould(2000)]{Gould2000} Gould, A. 2000, \apj, 542, 785
\bibitem[Gould \& Loeb(1992)]{Gould1992}  Gould, A., \& Loeb, A. 1992, \apj, 396, 104
\bibitem[Gould et al.(2006)]{Gould2006} Gould, A., Udalski, A., An, D., et al. 2006, \apj, 644, L37
\bibitem[Gould et al.(2022)]{Gould2022} Gould, A., Han, C., Zang, W., et al. 2022, \aap, 664, A13
\bibitem[Han et al.(2013)]{Han2013} Han, C., Jung, Y. K., Udalski, A. 2013, \apj, 778, 38
\bibitem[Han et al.(2020)]{Han2020} Han, C., Shin, I.-G., Jung, Y. K., et al. 2020, \aap, 641, A105
\bibitem[Han et al.(2022)]{Han2022} Han, C., Kim, D., Gould, A., et al. 2022, \aap, 664, A33
\bibitem[Herrera-Mart\'in et al.(2020)]{Herrera2020} Herrera-Mart\'in, A., Albrow, M. D., Udalski, A. et al. 2020, \aj, 159, 256
\bibitem[Hwang et al.(2022)]{Hwang2022} Hwang, K.-H., Zang, W., Gould, A., et al. 2022, \aj, 163, 43
\bibitem[Jung et al.(2018)]{Jung2018} Jung, Y. K., Udalski, A., Gould, A., et al. 2018, \aj, 155, 219
\bibitem[Jung et al.(2021)]{Jung2021} Jung, Y. K., Han, C., Udalski, A., et al. 2021, \aj, 161, 293
\bibitem[Jung et al.(2022)]{Jung2022} Jung, Y. K., Zang, W., Han, C., et al. 2022, \aj, 164, 262
\bibitem[Jung et al.(2024)]{Jung2024} Jung, Y. K., Hwang, K.-H., Yang, H., et al. 2024, \aj, 168, 152 
\bibitem[Kervella et al.(2004)]{Kervella2004} Kervella, P., Th\'evenin, F., Di Folco, E., \& S\'egransan, D.\ 2004, \aap, 426, 29
\bibitem[Kim et al.(2016)]{Kim2016} Kim, S.-L., Lee, C.-U., Park, B.-G., et al.\ 2016, JKAS, 49, 37
\bibitem[Kim et al.(2018)]{Kim2018} Kim, D.-J., Kim, H.-W., Hwang, K. -H. 2018, \aj, 155, 76
\bibitem[Kim et al.(2021a)]{Kim2021a} Kim, H.-W., Hwang, K.-H., Gould, A., et al. 2021a, \aj, 162, 15
\bibitem[Kim et al.(2021b)]{Kim2021b} Kim, Y. H., Chung, S.-J., Yee, J. C., et al. 2021b, \aj, 162, 17
\bibitem[Koshimoto et al.(2023)]{Koshimoto2023} Koshimoto, N., Sumi, T., Bennett, D. P., et al. 2023, \aj, 166, 107      
\bibitem[Mao \& Paczy\'nski(1991)]{Mao1991} Mao, S. \& Paczy\'nski, B. 1991, \apj, 374, L37
\bibitem[Matsuo et al.(2007)]{Matsuo2007} Matsuo, T., Shibai, H., Ootsubo, T., \& Tamura, M. 2007, \apj, 662, 1282
\bibitem[Miyazaki et al.(2018)]{Miyazaki2018} Miyazaki, S., Sumi, T., Bennett, D. P., et al. 2018, \aj, 156, 136
\bibitem[Mr\'oz et al.(2018)]{Mroz2018} Mr\'oz, P., Ryu, Y.-H., Skowron, J., et al. 2018, \aj, 155, 121
\bibitem[Mr\'oz et al.(2019)]{Mroz2019} Mr\'oz, P., Udalski, A., Bennett, D. P., et al. 2019, \aap, 622, A201
\bibitem[Mr\'oz et al.(2020a)]{Mroz2020a} Mr\'oz, P., Poleski, R., Gould, A., et al. 2020a, \apjl, 903, L11
\bibitem[Mr\'oz et al.(2020b)]{Mroz2020b} Mr\'oz, P., Poleski, R., Han, C., et al. 2020b, \aj, 159, 262
\bibitem[Nataf et al.(2013)]{Nataf2013} Nataf, D.~M., Gould, A., Fouqu\'e, P., et al.\ 2013, \apj, 769, 88
% -------
\bibitem[Ryu et al.(2021a)]{Ryu2021a} Ryu, Y.-H., Mr\'oz, P., Gould, A., et al. 2021a, \aj, 161, 126
\bibitem[Ryu et al.(2021b)]{Ryu2021b} Ryu, Y.-H., Hwang, K.-H., Gould, A., et al. 2021b, \aj, 162, 96
% -------
\bibitem[Shvartzvald et al.(2017)]{Shvartzvald2017} Shvartzvald, Y., Yee, J. C., Novati, S. C., et al. 2017, \apjl, 840, L3
\bibitem[Sumi et al.(2016)]{Sumi2016} Sumi, T., Udalski, A., Bennett, D. P., et al. 2016, \apj, 825, 112
\bibitem[Suzuki et al.(2016)]{Suzuki2016} Suzuki, D., Bennett, D. P., Sumi, T., et al. 2016, \apj, 833, 145
\bibitem[Tsapras et al.(2003)]{Tsapras2003} Tsapras, Y., Horne, K., Kane, S., \& Carson, R. 2003, \mnras, 343, 1131
\bibitem[Udalski et al.(1994)]{Udalski1994} Udalski, A., Szymanski, M., Ka{\l}u\.zny, J., et al. 1994, Acta Astron., 44, 1
\bibitem[Yang et al.(2024)]{Yang2024} Yang, H., Yee, J. C., Hwang, K.-H., et al. 2024, \mnras, 528, 11
\bibitem[Yee et al.(2012)]{Yee2012} Yee, J. C., Shvartzvald, Y., Gal-Yam, A., et al. 2012, \apj, 755, 102
\bibitem[Yee et al.(2021)]{Yee2021} Yee, J. C., Zang, W., Udalski, A., et al. 2021, \aj, 162, 180
\bibitem[Yoo et al.(2004)]{Yoo2004} Yoo, J., DePoy, D. L., Gal-Yam, A. et al. 2004, \apj, 603, 139
\bibitem[Zhang et al.(2022a)]{Zhang2022a} Zhang, K., Gaudi, B. S., \& Bloom, J. S. 2022, Nat Astron, 6, 782 
\bibitem[Zhang \& Gaudi(2022b)]{Zhang2022b} Zhang, K., \& Gaudi, B. S. 2022, \apjl, 936, L22 
\bibitem[Zhang et al.(2020)]{Zhang2020} Zhang, X., Zang, W., Udalski, A., et al. 2020, \aj, 159, 116
%%====================================

%\vspace*{\fill}
\end{thebibliography}
\end{document}